\documentclass{sig-alternate}
  \usepackage{color}

\newcommand{\tocheck}[1]{#1}

\usepackage{helvet}

\usepackage{courier}
\usepackage{booktabs}
\usepackage{tabularx}
\usepackage{amssymb}
\usepackage{graphicx}
\usepackage{lipsum}
\usepackage{epsfig}
\usepackage{amsmath}
\usepackage{multirow}
\usepackage{subcaption}
\usepackage[para]{footmisc}
\usepackage{url}
\usepackage{tikz}
\usepackage{enumitem}
\usepackage{amsfonts}
\usepackage{multicol}
\usepackage{color,soul}

\usepackage{makeidx}
\usepackage{graphicx}
\usepackage{amsmath}
\usepackage{amsfonts}
\usepackage{amssymb}
\usepackage{epsfig}
\usepackage{epstopdf}
\usepackage{verbatim}
\usepackage{tabularx}
\usepackage{color}
\usepackage{times}
\usepackage{url}
\usepackage{latexsym}
\usepackage{float}

\usepackage{pgfplots}

\def\sharedaffiliation{%
\end{tabular}
\begin{tabular}{c}}

\begin{document}

\numberofauthors{3} 
\author{
\alignauthor
Jarana Manotumruksa$^1$\\
\alignauthor
Craig Macdonald$^2$\\
\alignauthor 
Iadh Ounis$^2$\\
\sharedaffiliation\vspace{-1\baselineskip}
       \affaddr{University of Glasgow, UK}\\
	\email{$^1$j.manotumruksa.1@research.gla.ac.uk}, 
    \email{$^2$\{firstname.lastname\}@glasgow.ac.uk}
}

%
\conferenceinfo{Neu-IR '16 SIGIR Workshop on Neural Information Retrieval,}{July 21, 2016, 
Pisa, Italy}
\CopyrightYear{2016} 

\CopyrightYear{2016}

\title{Modelling User Preferences using Word Embeddings for Context-Aware Venue Recommendation}

\maketitle
\begin{abstract}
\looseness -1 Venue recommendation aims to assist users by making personalised suggestions of venues to visit, building upon data available from location-based social networks (LBSNs) such as Foursquare. A particular challenge for this task is context-aware venue recommendation (CAVR), which additionally takes the surrounding context of the user (e.g.\ the user's location and the time of day) into account in order to provide  more relevant venue suggestions. To address the challenges of CAVR, we describe two approaches that exploit word embedding techniques to infer the vector-space representations of venues, users' existing preferences, and users' contextual preferences. Our evaluation upon the test collection of the TREC 2015 Contextual Suggestion track demonstrates that we can significantly enhance the effectiveness of a state-of-the-art venue recommendation approach, as well as produce context-aware recommendations that are at least as effective as the top TREC 2015 systems.



%
\end{abstract}

\section{Introduction}\label{sec:intro}
\enlargethispage{2\baselineskip}
Unlike traditional item recommender tasks, recommending relevant venues for a user to visit relies not only his/her venue preferences, but also on contextual information about the user, such as the user's location and time of day. Previous work on context-aware venue recommendation (CAVR) showed that by considering such contextual information, more relevant venue suggestions can be achieved~\cite{yao2015context,yuan2013time}. Recently, producing effective rankings of venues under more challenging contextual constraints such as {\em Season} (summer, winter, ...), travelling {\em Group} (alone, with friends, or with family) have been tackled by the research community~\cite{trec2015amsterdam,trec2015glasgow}, facilitated by the TREC 2015 Contextual Suggestion track~\cite{dean2015overview}. 

\looseness -1 Most of the literature on CAVR relies on collaborative filtering approaches. However, such approaches require a large training dataset to learn the number of model parameters, which grow exponentially as the number of contextual factors increases~\cite{karatzoglou2010multiverse}. In addition, by its very nature, CAVR is most useful for users visiting new cities (i.e.\ without previous venue preferences), which reduces the effectiveness of collaborative filtering approaches for this task. 

\looseness -1 Many existing approaches for venue recommendation make use of textual and statistical data about each venue, such as the contents of the venue's website, or comments posted by users about the venue in Location-based Social Networks (LBSNs) such as Foursquare\footnote{https://foursquare.com} and Yelp\footnote{http://www.yelp.com}, or the number of user check-ins\footnote{Foursquare terminology denoting users sharing their  location with the LBSN.}~\cite{trec2015lugano,deveaud2014importance,trec2015amsterdam,trec2015glasgow,yang2013opinion}. Textual data can be useful for representing the user profiles and venue profiles for the purposes of content-based collaborative filtering, but may be sparse due to the variance in words used by users to express their opinions about venues. Moreover, how best to match and rank venues according to  abstract contextual aspects such as \textit{Group} (e.g.\ What venues are suitable to visit for a family trip?) remains an open problem, due to the difficulty of finding terms that are related to contextual aspects within comments about the venues. 


\looseness -1 To address these challenges, we turn to recent research using word embeddings as a source of semantically related terms. In doing so, this paper makes two contributions: (1) we adapt a recent content-based collaborative filtering approach based on word embeddings~\cite{musto1441word} to effectively identify venues that match the users' {\em venue} preferences, and (2) building upon this, we propose a novel approach that models the users' {\em contextual} preferences using word embeddings to ensure accurate venue suggestions. Our experiments, conducted within a learning to rank framework and evaluated using the TREC 2015 Contextual Suggestion track test collection, demonstrate that both word embedding applications can significantly enhance the recommendation effectiveness, outperforming state-of-the-art-approaches from the literature and TREC 2015.

\section{Related Work}\label{sec:related}
\enlargethispage{2\baselineskip}
\looseness -1 Venue recommendation can be tackled using collaborative filtering techniques, taking into account various types of context. For instance, Yuan {\em et al.}~\cite{yuan2013time} extended a collaborative filtering technique to incorporate temporal information about the preferences of users for venues, in order to generate venue recommendations that are aware of the temporal context of users. Many previous works on CAVR~\cite{ozsoy2016word,yao2015context,yuan2013time} use check-in data from LBSNs to evaluate the accuracy of their recommendations, by assuming that users implicitly like the venues they visited. A more reliable and reusable test collection methodology to evaluate CAVR systems has been developed by the TREC Contextual Suggestion track~\cite{dean2013overview} since 2012. In particular, the task addressed by the track is as follows: given the user's preferences (ratings of venues) and context (e.g.\ user's location, city), to produce a ranked list of venue suggestions for each user-context pair. However, while there can be various different types of contextual information associated with a user, the approach of Yuan {\em et al.}~\cite{yuan2013time} can only encapsulate time-based context. The tensor factorisation technique proposed by Yao {\em et al.}~\cite{yao2015context} is more extensible as it can support multi-dimensional contextual information, but it requires a large dataset to train the model's parameters. This is difficult to attain in CAVR, where users may visit a venue in varying contextual situations, or may have never visited a city before. 


\looseness -1 Many of the approaches tested in the TREC tracks have relied upon LBSNs to obtain user-generated data about venues such as a number of check-ins by users or comments users have posted about venues~\cite{deveaud2014importance,mccreadie2013university,yalamartiyork,yang2013opinion}. For instance, Yang {\em et al.}~\cite{yang2013opinion} proposed an approach that constructs positive and negative user profiles from the venue's comment that users rated positively (e.g.\ four stars) and negatively (one star), respectively. The venue profiles are constructed the same way as the user profiles, such that candidate venues can be ranked by the cosine similarity score between the user and venue profiles. The intuition behind this approach -- which was shown to be effective in the TREC 2013 and 2014 Contextual Suggestion tracks -- is that people who have similar opinions about venues are likely to share similar venue preferences. Likewise, our work builds upon that of Yang {\em et al.}~\cite{yang2013opinion}, but we propose to model the user's preferences by using word embeddings to better represent the venues as vectors in the word embedding space.




\looseness -1 Word embeddings have recently gained a lot of attention due to their effectiveness in capturing linguistic regularities for text processing problems. Indeed, within a recommendation task, Musto {\em et al.}~\cite{musto1441word} conducted an empirical study to evaluate the effectiveness of a simplified collaborative filtering recommendation framework that applies several word embedding techniques to model items and make recommendations to users. In particular, they mapped the items (movies and books) in their datasets to textual content using Wikipedia and used terms in the textual content to model the items using word embeddings. Ozsoy {\em et al.}~\cite{ozsoy2016word} adapted the skip-gram technique proposed by Mikolov {\em et al.}~\cite{mikolov2013efficient,mikolov2013linguistic} to model users and venues in a vector space and made predictions on where the users will check-in next. In contrast to these previous works, we propose two approaches that exploit word embeddings to extract both discriminative user-venue and context-venue preference features based on user-generated comments posted about the venues on a LBSN. Our experimental results show not only the promise of word embeddings for venue recommendation, but also demonstrate significant improvements over a state-of-the-art user modelling approach~\cite{yang2013opinion}.

\section{Problem Statement}\label{sec:prob}
\enlargethispage{2\baselineskip}
\looseness -1 We now formally describe the notations and the context-aware venue recommendation task of the TREC 2015 Contextual Suggestion track~\cite{dean2015overview}. First, the {\em recommendation of venues} for a user can be expressed as follows: Let $V$ be a set of venues $ \left\{ v_1, \ldots, v_i \right\}$ and $U$ be a set of users $ \left\{ u_1, \ldots, u_j \right\}$. Each user $u_j$ has expressed preferences about some venues, denoted as $ U_j = \left\{v_i \rightarrow r_{i,j}, \mbox{ } \ldots \right\}$, where $r_{i,j}$ is a preference rating (1 to 5) of user $u_j$ for venue $v_i$. Additionally, $u_{j,gender}$ indicates the gender of user $u_j$ (male or female). Therefore, the task is to estimate whether a venue $v_i$ is appropriate for user $u_j$ to visit.

\begin{table}[tb]
\centering\scriptsize
    \begin{tabular}{|l|l|}
    \hline
    \textbf{Aspect} & \textbf{Applicable contextual dimensions}                                                           \\ \hline
    Duration                    & Day Time, Night Time and Weekend       \\ \hline
    Season                    & Spring, Summer, Autumn and Winter       \\ \hline
    Group                    & Alone, with Friends and with Family       \\ \hline
    Type                    & Business and Holiday trip      \\ \hline
    \end{tabular}
\caption{Contextual dimensions \& aspects considered in this work.}\label{tab:description}\vspace{-4mm}
\end{table}


\looseness -1 A context-aware venue recommendation task~\cite{dean2015overview} requires that the suggested venues are appropriate for explicit contextual requirements of the user. Let $A$ be a set of contextual aspects that the users can express. Each aspect $a \in A$ has a set of dimensions, $ a = \left\{ d_{a,1} \ldots d_{a,k} \right\} $. A summary of dimensions and aspects is shown in Table~\ref{tab:description}. Therefore, the task in the TREC Contextual Suggestion track is to rank a list of venues in response to a user-context pair, denoted $\langle U_j, C_j \rangle$, where $C_j$ is a set of contextual dimensions applicable to user $j$: $ C_j = \left\{ d, \mbox{ } \ldots \right\}$. Note that only one dimension can be expressed for each of the aspects, $|C_j| = |A| \leq 4$.

\section{CAVR using Word Embeddings}\label{sec:method}
\looseness-1 We now describe two approaches to improve venue recommendations that use word embeddings in the modelling of the user-venue (Section~\ref{sec:method:uv}) \& context-venue preferences (Section~\ref{sec:method:uc}).

\subsection{User-Venue Preferences}\label{sec:method:uv}
Similar to the previous work of Musto et al.~\cite{musto1441word} on recommending movies and books, we exploit a word embedding technique to infer a vector-space representation of a venue. In particular, as we use a LBSN to obtain data about venues, each venue $v_i$ has an associated set of comment posts $P_i = \{p_{i,1}, ..\}$. 
A representation of the venue within the word embedding space, denoted $W2V(v_i)$, is obtained by summing the vector $\vec{\nu}_t$ of each term $t$ occurring in the venue's comments $P_i$:
\begin{equation}
W2V(v_i) = \sum_{p \in P_i} \sum_{t \in p} \vec{\nu}_t
\end{equation}
Similarly, we model the user-venue preferences, $\vec{UV_j}$, of user $u_j$ by summing the vector of the venues rated in the user's profile $U_j$:
\begin{equation}
\vec{UV}_j = \sum_{v_i \in U_j} W2V(v_i) \times r_{i,j}
\end{equation}

\looseness -1 We then estimate the similarity between venue $v_i$ and the user-venue preferences $\vec{UV}_j$ by calculating the cosine similarity between $\vec{UV}_j$ and $W2V(v_i)$. Typically, the user-venue preferences $\vec{UV}$ are modelled separately for positive (rating of 3 or 4) vs.\ negative (0 or 1) rated venues in the user's profile.


\enlargethispage{2\baselineskip}
\subsection{Context-Venue Preferences}\label{sec:method:uc}
Building upon $W2V(v_i)$ defined above, we now propose a novel approach for CAVR that models how well a venue matches the user's contextual preferences, $\vec{CV}$. A key property of word embeddings is that semantically related terms exhibit low distances in the vector space. We exploit this property to identify a list of terms that are related to each contextual dimension. For example, terms that are close to the word `Family' in the vector space are ``grandparents'', ``supporting'', ``group'', ``mom'', ``kid'', etc. Observing such words frequently in the comments posted for a venue on a LBSN intuitively increases our belief in that venue being relevant for recommendation scenarios involving \textit{Family} contextual requirements.

Hence, for a given contextual dimension $d_{a,k}$, we aim to find a textual representation that can be matched to the comments of a venue. To achieve this, we take the name of dimension (e.g.\ Family), and identify the $K$ most similar terms, $Similar_K(\vec{\nu}_{d_{a,k}})$, according to the word embedding space.




However, the various dimensions of the same contextual aspect are likely to share related terms (e.g.\ the comments about a venue suitable to visit in Spring and Autumn may contain similar terms). Table~\ref{tab:illu} illustrates a list of related terms for each of the contextual dimensions in the \textit{Season} aspect. Intuitively, the terms that are bold in Table~\ref{tab:illu} are not related to a contextual dimension term, although they are close in the vector space. These terms might result in a less effective modelling of the user's contextual preferences.

\begin{table}[tb]
\resizebox{82mm}{!}{
    \begin{tabular}{|l|l|}
    \hline
    Contextual Dimension term & A list of related terms                                                            \\ \hline
    Spring                    & vermicelli, \textbf{summer}, pho, rolls, pad, kha, rangoon, cuon         \\ \hline
    Summer                    & \textbf{winter}, summertime, \textbf{spring}, warmer, rainy, fall, chilly   \\ \hline
    Autumn                    & \textbf{winter}, pumpkin, fall, squash, mulled, harvest, \textbf{snowy} \\ \hline
    Winter                    & chilly, \textbf{warmer}, \textbf{summer}, rainy, weather, freezing, fall     \\ \hline
    \end{tabular}}
\caption{Related terms for the dimensions of the Season aspect.}\label{tab:illu}\vspace{-4mm}
\end{table}

\looseness -1 It has been shown that the word vectors capture many linguistic regularities~\cite{mikolov2013efficient,mikolov2013linguistic}. For instance, applying subtraction operations on two vectors, $\vec{\nu}_{summer} - \vec{\nu}_{winter}$ results in vectors that are close to $\vec{\nu}_{summer}$ but not close to $\vec{\nu}_{winter}$ in the vector space. Hence, to obtain a list of terms that are related to only one contextual dimension of each aspect, we propose to aggregate the terms obtained from a subtraction operation of that contextual dimension over other dimensions that are in the same contextual aspect, i.e.\ the set of terms $T_{d_{a,k}}$ for dimension $d_{a,k}$ are obtained as:
\begin{equation}
T_{d_{a,k}} = \bigcup_{d_{a,i} \in a \wedge d_{a,i} \neq d_{a,k}} Similar_K(\vec{\nu}_{d_{a,k}} - \vec{\nu}_{d_{a,i}})
\end{equation}
\looseness -1 Finally, we model a user's contextual preferences $\vec{CV}$ for dimension $d_{a,k}$ as follows:
\begin{equation}
\vec{CV}_{d_{a,k}} = \sum_{t \in T_{d_{a,k}}} \vec{\nu}_t
\end{equation}
Therefore, to obtain the final score of a venue for a contextual dimension $d_{a,k}$ preference expressed by a user (i.e.\ $d_{a,k} \in C_j$), we calculate the cosine similarity in the word embedding space between $\vec{CV}_{d_{a,k}}$ and the comments of the venue, i.e.\ $W2V(v_i)$. \tocheck{Note that later in Section~\ref{ssec:expContext} we show how our approach can be extended to also encapsulate the gender-venue preferences of the user.}


\section{Evaluation}\label{sec:eval}
\enlargethispage{2\baselineskip}
\looseness-1 In this section, we evaluate our two proposed approaches, comparing with top-performing approaches participating in the TREC Con\-t\-ext\-ual Suggestion tracks. In particular, we address two research questions: \textbf{RQ1} Can we exploit word embeddings to model user-venue preferences to improve the quality of venue suggestions? (Section~\ref{sec:eval:uv}); \textbf{RQ2} Can {\em context}-venue preferences be modelled using word embeddings to attain effective CAVR? (Section~\ref{sec:eval:uc}).

\subsection{Modelling User-Venue Preferences}\label{sec:eval:uv}
\looseness -1 To answer \textbf{RQ1}, we compare our model for user-venue preferences $\vec{UV}$ that uses word embeddings (Section~\ref{sec:method:uv}) with a state-of-the-art user modelling approach~\cite{yang2013opinion} that attained top performances in the TREC 2013 \& 2014 Contextual Suggestion tracks~\cite{dean2013overview,dean2014overview}.


\subsubsection{Experimental Setup}\label{sssec:setup1}
\looseness-1 We use the test collection of the TREC 2015 Contextual Suggestion track, which provides 211 user-context pairs, $\langle U_j, C_j \rangle$, with venue relevance assessments by those users. Note that in this experiment, for each user-context pair, we only consider the users' profiles, $U_j$, as we aim to evaluate the effectiveness of our proposed approach that models user-venue preferences $\vec{UV}$. Later in Section~\ref{ssec:expContext}, we will evaluate the effectiveness of a CAVR system that additionally considers the users preferred contexts $C_j$ and their genders $u_{j,gender}$ when ranking venues.

\looseness -1 We use three learning to rank techniques, namely Coordinate Ascent -- which creates a linear combination of features -- from the RankLib toolkit, as well as two techniques based on boosted regression trees as implemented by the Jforests toolkit: MART -- a pointwise technique -- and the state-of-the-art listwise LambdaMART~\cite{Wu08ranking}. Each learning technique is trained on features extracted from our proposed approach or the baseline approach~\cite{yang2013opinion}. In particular, each venue is represented using features extracted from its corresponding comments on Four\-square (called tips). We calculate two different set of features as follows:

\looseness -1 \noindent\textbf{Our Approach (denoted as Ours)} - 2 user-venue preference features: The cosine similarity between the vector-space representation of venue $W2V(v_i)$ and the vector-space representation of the user-venue preferences $\vec{UV}_j$ of user $u_j$. One feature for positive user-venue preferences, and one for negative preferences.

\looseness -1 \noindent\textbf{Baseline} - 4 rating-based features: Cosine similarity scores computed between the positive and negative user profiles and each venue, which are extracted from the comments posted about venues~\cite{yang2013opinion}.


All learners are trained on the 60 venue preferences $U_j$ of all users provided by the TREC track-- as these are separate from the test venues. This represents a clear separation between training and test environments. We use 67\% of the venue preferences for training each learner, and 33\% for validation (e.g.\ setting the number of learning iterations).

\looseness -1 For word embeddings, we use the Word2Vec tool\footnote{https://code.google.com/archive/p/word2vec/}, training a skip-gram model~\cite{mikolov2013efficient} by using default settings (vector size = 100 and window size = 5) on the venues' comments dataset from Foursquare. Before training the model, we perform stemming and remove stopwords from the comments dataset. We choose the skip-gram model over the continuous bag of word (CBOW) as it is reported to perform better than or equally to the alternative CBOW technique~\cite{mikolov2013distributed}. Finally, following~\cite{dean2015overview}, we report effectiveness in terms of Precision@5 (P@5) and mean reciprocal rank (MRR).


\subsubsection{Experimental Results}
\label{sssec:result1}

\begin{table}[tb]
\centering
\resizebox{85mm}{!}{
\begin{tabular}{|l|c|c|c|c|c|c|}
\hline
                   & \multicolumn{2}{c|}{P@5} & $\Delta$   & \multicolumn{2}{c|}{MRR} & $\Delta$   \\ \hline
Model              & Baseline     & Ours      &         & Baseline     & Ours      &         \\ \hline
Coordinate Ascent           & 0.5242       & 0.5773    & 9.2\%***  & 0.6531       & 0.7008    & 6.8\%*    \\ \hline
MART       & 0.4853       & \textbf{0.5791 }   & 16.2\%*** & 0.6390       & \textbf{0.7285}    & 12.3\%*** \\ \hline
LambdaMART & 0.5213       & 0.5630    & 7.4\%**   & 0.6575       & 0.7059    & 6.9\%*    \\ \hline

\end{tabular}
}
\caption{\looseness -1 Effectiveness results for the different learning to rank models. $\Delta$ denotes the relative improvement of our proposed approach over the baseline. Significant improvements are indicated by *, ** and *** according to paired t-test (p < 0.05, 0.01 and 0.001 resp.).}
\label{tab:uservenue}\vspace{-3mm}
\end{table}

\looseness-1 To address RQ1, Table~\ref{tab:uservenue} shows the effectiveness results of the learned models trained on features extracted from our proposed approach or Yang's approach~\cite{yang2013opinion}. From the table, we observe that all the models trained on \textbf{Ours} features outperform the models trained on the \textbf{Baseline} features, for both P@5 and MRR. For instance, significant improvements of 7-9\% can be obtained over the performances of the strongest learners LambdaMART and Coordinate Ascent (paired t-test). These results show the effectiveness of representing user-venue preferences using word embeddings.





\enlargethispage{2\baselineskip}
\subsection{Context-Aware Recommendations}\label{sec:eval:uc}
\label{ssec:expContext}
To address RQ2, we deploy a CAVR system that generates personalised context-based venue suggestions, by encapsulating the context-venue approach described in Section~\ref{sec:method:uc}, as well as the features described in Section~\ref{sec:method:uv} within a learning-to-rank context. In particular, we deploy a total of 13 venue ranking features, building upon the MART learner (the best learner in Section~\ref{sec:eval:uv}), and compare the resulting system with the top-performing approaches in the TREC 2015 Contextual Suggestion track.


\noindent\textbf{6 venue-dependent features}: Number of check-ins, number of likes, number of comments, number of photos, average Four\-square rating, and unique number of users~\cite{deveaud2014importance}.
 
\looseness -1 \noindent \textbf{2 user-venue preference features, aka Ours}: As in Section~\ref{sec:eval:uv}, Cosine similarity between the vector-space representation of venue $W2V(v_i)$ and the vector-space representation of the user-venue preference $\vec{UV}_j$ of user $u_j$ -- one feature each for positively ($r_ij$$\geq$4) and negatively ($r_ij$$\leq$3) rated venues preferences (denoted UV\_pos \& UV\_neg, resp.).

\noindent\looseness -1\textbf{4 context-venue preference features}: Cosine similarity between the vector-space representation of venue $W2V(v_i)$ and user's contextual preference $\vec{CV}_{j, d_{a,k}}$, where $d_{a,k} \in C_j$ -- one feature for each contextual aspect, as proposed in Section~\ref{sec:method:uc}.

\noindent\looseness -1 \textbf{1 gender-venue preference feature}: Cosine similarity between $W2V(v_i)$ and the gender-venue preference $\vec{GV}_j$, which is generated similarly to $\vec{CV}$, using $T_{u_{j,gender}}$.


Finally, the $K$ parameter, which defines the number of similar terms to consider for each contextual dimension (see Section~\ref{sec:method:uc}), is set using the 33\% validation data that are obtained from the user's venue preferences profile ($U_j$).

\subsubsection{Experimental Results}
\label{sssec:resultscavr}

\begin{table}[tb]
\resizebox{85mm}{!}{
\begin{tabular}{|c|c|c|c|c|c|c|}
\hline
      & User & Venue & Context& Gender & P@5 & MRR \\
\hline
\hline
All 13 & \checkmark  & \checkmark & \checkmark & \checkmark & \textbf{0.5972} & 0.7340 \\
6 venue-dependent  & $\times$  & \checkmark & $\times$ & $\times$ & 0.5668* (-5.1\%) & 0.6891* (-6.1\%) \\
\hline
\hline
TREC Median & - & - & - & - & 0.5090 & 0.6716 \\
\hline
USI & \checkmark  & \checkmark & $\times$ & $\times$ & 0.5858  & \textbf{0.7404} \\
uogTr & \checkmark  & \checkmark & \checkmark & $\times$ & 0.5706 & 0.7190 \\
UAmsterdam & \checkmark  & \checkmark & \checkmark & \checkmark & 0.5204 & 0.6765 \\
\hline
\end{tabular}
}
\caption{Results of the top-performing TREC 2015 Contextual Suggestion track systems compared with our proposed approach trained on 13 features (denoted as All). The best performances for each measure are emphasised, while performances denoted * exhibit significant decreases in effectiveness (paired t-test, $p<0.05$) compared to the approach trained on All features}\label{tab:trecresults}\vspace{-4mm}
\end{table}

\looseness -1 The top part of Table~\ref{tab:trecresults} reports the effectiveness of a MART model based on all 13 features, as well as a model applying only the 6 venue dependent features. Indeed, this latter model forms a baseline, in that it does not consider any evidence about the user or their contextual preferences. The high effectiveness of the venue-dependent features has previously been reported in ~\cite{deveaud2014importance}. Comparing these two learned models, we note that the 7 user- and context-venue features provide statistically significant improvements in both P@5 and MRR (5-6\%, $p<0.05$, paired t-test).

\looseness -1 The second part of Table~\ref{tab:trecresults} provides comparative performances from the TREC 2015 Contextual Suggestion track: the per-topic median, and 2 top-performing groups (USI~\cite{trec2015lugano} \& uogTr~\cite{trec2015glasgow}), as well as the UAmsterdam group~\cite{trec2015amsterdam}. In particular, we select the latter as their run encapsulates the same sources of evidence as our approach, namely the user-venue interestingness (denoted as User), the venue information (Venue), the contextual sources of evidence (Context) and the user's Gender. Our MART model with 13 features outperforms all submitted TREC 2015 runs in terms of P@5, and is as effective as the top-performing approach (USI) for MRR.\footnote{As the actual rankings of the participating systems are not yet available, it is not possible to perform significance testing between our approach and these systems.}

\begin{figure}[tb]
\begin{tikzpicture}
\begin{axis}[
	xticklabels from table = {ablateExp.dat}{Num},
    x tick label style={rotate=45, anchor=east},
	ylabel= $\Delta$ \% ,
	legend style={at={(0.5,1.2)},
	anchor=north,legend columns=-1},
    x=0.8cm, y=0.6cm/1.5,
        ymajorgrids = true,
	ybar interval=0.7]
    \addplot table [x expr = \lineno, y = p5] {ablateExp.dat};

\end{axis}
\end{tikzpicture}\vspace{-4.5mm}
\caption{Relative change in P@5 ablating single features from the MART model that uses all 13 features.}
\label{fig:features}\vspace{-5mm}
\end{figure}
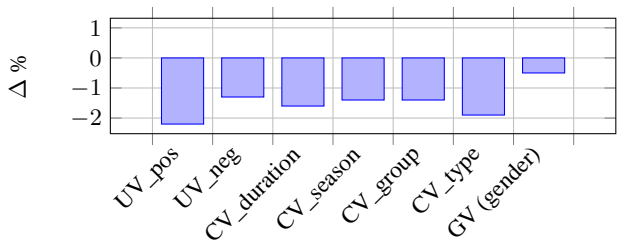

Finally, to determine the importance of the proposed features, we additionally perform a feature ablation test, by recording the effectiveness when a single feature is removed from the total 13\footnote{We omit the ablation of the venue-dependent features, which is already studied in~\cite{deveaud2014importance}.}. The relative \% change in P@5 is recorded in Figure~\ref{fig:features}, where the most useful feature records the lowest $\Delta$, (-2.2\% for UV\_pos). Indeed, on analysing the figure, we note that abalting the user-venue preference feature based on positive venue ratings (UV\_pos) has the largest negative impact on P@5 effectiveness. Each of the context-venue preference features reduce P@5 by 1.4-1.9\%. These results suggest that the context-venue preferences based on word embeddings are effective, but represent factors that are less important for users than their previous venue preferences. Lastly, ablation of the gender-venue feature only reduces effectiveness by -0.5\%.

\looseness -1 Overall, for RQ2, we find that our user-venue and context-venue features based upon word embeddings can significantly enhance a competitive learned venue recommendation approach, and outperform the top TREC 2015 participating system under P@5.

\section{Conclusions}\label{sec:con}
\enlargethispage{\baselineskip}
\looseness-1 Although the task of CAVR can be addressed using collaborative filtering approaches, by their very nature such approaches suffer from a data sparsity problem. In this work, we examined two approaches that model user-venue and context-venue preferences in terms of a word embedding vector-space. Experiments upon the TREC 2015 Contextual Suggest track demonstrated the effectiveness of our proposed approaches in comparison with state-of-the-art-approaches from both the literature and TREC 2015.



\small
\bibliography{foo}
\bibliographystyle{abbrv}

\end{document}